\documentclass[prb,twocolumn,amssymb,showpacs]{revtex4}
\usepackage{graphicx}
\usepackage{amsmath}

\begin{document}

\title{Magnetic order and spin fluctuations in the spin liquid Tb$_2$Sn$_2$O$_7$.}
\author{I. Mirebeau$^1$, A. Apetrei$^1$, J. Rodr\'{\i}guez-Carvajal$^1$, P. Bonville$^2$, A. Forget$^2$, D. Colson$^2$,
V. Glazkov$^3$, J. P. Sanchez$^3$, O. Isnard$^{4,5}$ and E.
Suard$^5$.
\\
}
\address{
$^1$Laboratoire L\'eon Brillouin, CEA-CNRS, CE-Saclay, 91191
Gif-sur-Yvette, France.}
\address {$^2$Service de Physique de l'Etat Condens\'e,
CEA-CNRS, CE-Saclay,  91191 Gif-Sur-Yvette, France.}
\address {$^3$Service de
Physique Statistique, Magn\'etisme et Supraconductivit\'e,
CEA-Grenoble, 38054 Grenoble, France.}
\address{$^4$Laboratoire de Cristallographie, Univ. J. Fourier-CNRS, BP 166, 38042 Grenoble France.}
\address{$^5$Institut La\"ue Langevin, 6 rue Jules Horowitz, BP 156X, 38042
Grenoble France.}

\begin{abstract}
We have studied the spin liquid Tb$_2$Sn$_2$O$_7$ by
 neutron diffraction and specific heat measurements.
Below about 2 K, the magnetic correlations change from
antiferromagnetic
 to ferromagnetic. Magnetic order settles in two steps, with a smeared transition
at 1.3(1) K then an abrupt transition at 0.87(2) K. A new magnetic
structure is observed, not predicted by current models, with both
ferromagnetic and antiferromagnetic character. It suggests that
the spin liquid degeneracy is lifted by dipolar interactions
combined with a finite anisotropy along $<$111$>$ axes.
  In the ground state, the Tb$^{3+}$ ordered moment is reduced with respect to the free ion
  moment (9 $\mu_{\rm B}$). The moment value of 3.3(3) $\mu_{\rm B}$ deduced from the specific heat is much smaller than derived from
neutron diffraction of 5.9(1) $\mu_{\rm B}$. This difference is
interpreted by the persistence of slow collective magnetic
fluctuations down to the lowest temperatures.

\end{abstract}

 \maketitle

Geometrically frustrated  pyrochlores  R$_2$Ti$_2$O$_7$ show
exotic magnetic behaviors
\cite{Harris97,Gardner99,Hodges02,Champion0301}
such as 
 dipolar spin ice (R=Dy, Ho) and spin liquid (Tb) phases, a first order transition in the spin dynamics (Yb),
 or 
complex antiferromagnetic orders (Er, Gd). The type of magnetic
order depends on the balance between antiferromagnetic exchange,
dipolar and crystal
field energies \cite{Siddharthan99,DenHertog00}. 
Tb$_2$Ti$_2$O$_7$ is a unique case of a spin liquid where short-
ranged correlated magnetic moments fluctuate down to 70 mK, with
typical energies 300 times lower than the energy scale given by
the Curie-Weiss constant $\theta_{\rm CW}$ of -19 K. The fact that
Tb$_2$Ti$_2$O$_7$ does not order at ambient pressure
\cite{Gardner99}, but could order under applied pressure, stress,
and magnetic field \cite{Mirebeau02, Mirebeau04}, is still a
challenge to theory, since recent models predict  a transition
towards antiferromagnetic (AF) long range order at about 1
K\cite{Kao03, Enjalran04}.

 With respect to titanium, substitution by tin yields a lattice
 expansion. It also modifies the oxygen environment of the Tb$^{3+}$ ion and therefore the crystal field.
 The stannates R$_2$Sn$_2$O$_7$ show the same crystal structure\cite{Kennedy97} as
the titanates, and susceptibility data\cite{Bondah01,Matsuhira02}
also suggest a great variety of magnetic behaviors.
Dy$_2$Sn$_2$O$_7$ and Ho$_2$Sn$_2$O$_7$ are dipolar spin
ices\cite{Matsuhira00,Kadowaki02} like their Ti parent compounds,
whereas Er$_2$Sn$_2$O$_7$ does not order down to 0.15
K\cite{Matsuhira02}, and Gd$_2$Sn$_2$O$_7$ undergoes a transition
to AF order\cite{Bonville03}. In Tb$_2$Sn$_2$O$_7$, magnetic
measurements suggest an original and complex behavior. 
Antiferromagnetic interactions are observed at high temperature,
yielding a Curie-Weiss constant $\theta_{\rm CW}$ of -11 to -12 K
\cite{Bondah01,Matsuhira02}, but a ferromagnetic (F) transition is
seen around 0.87 K \cite{Matsuhira02}.

We have performed neutron diffraction and specific heat 
 measurements
 in Tb$_2$Sn$_2$O$_7$. With decreasing temperature a spin liquid phase is shown to transform into a
 new type of ordered phase, not predicted by theory,
 which could be called an "ordered spin ice".
 Just above the transition, an abnormal change in the spin
 correlations shows
 the influence of dipolar interactions. By comparing the ordered Tb$^{3+}$
 moment values from neutron diffraction and nuclear specific heat, we also indirectly observe
 slow fluctuations of correlated spins, which persist 
 down to the lowest temperature.

  A powder sample of Tb$_2$Sn$_2$O$_7$ was synthesized. 
  The crystal
  structure with space group Fd$\overline{3}$m
   was studied at 300 K by combining X-ray diffraction
   with a neutron diffraction pattern measured in the
    diffractometer 3T2 of the Laboratoire L\'eon Brillouin (LLB). 
     Rietveld refinements performed with 
      Fullprof\cite{Carvajal93} confirmed the structural model
      (R$_{\rm B}$=2.4 \%), yielding
 the lattice constant
    a = 10.426 $\AA$ and
 oxygen
position parameter x=0.336. 
 The magnetic diffraction patterns were recorded between 1.4 K and 300 K and down to 0.1 K in the
 diffractometer G6-1 (LLB) and D1B of the Institut La\"ue Langevin (ILL) respectively.
  The specific heat
was measured by the dynamic adiabatic method down to 0.15 K.

\begin{figure} [h]
\includegraphics* [width=\columnwidth] {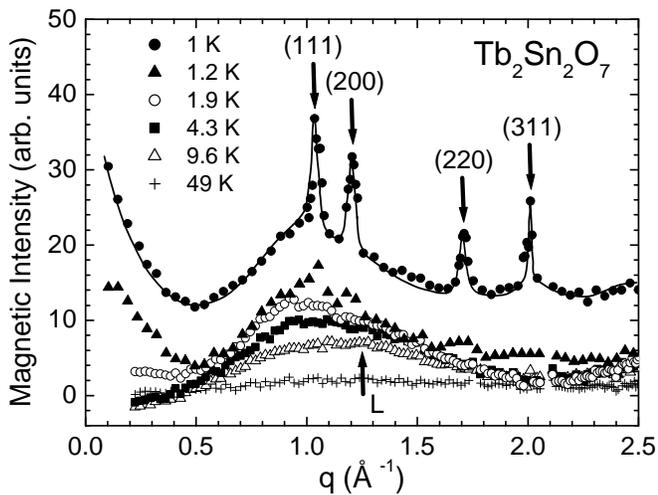}
\caption{Magnetic intensity of Tb$_2$Sn$_2$O$_7$ versus the
scattering vector
 q=4$\pi$sin$\theta$/$\lambda$. A spectrum in the paramagnetic phase (100 K) was subtracted. Intensities at 1. K have an offset of 10 for clarity. 
 Arrows show the position of the Bragg peaks and near neighbor liquid peak (L)
 calculated in ref\cite{Canals01}.
} \label{fig1.eps}
\end{figure}

 Fig. 1 shows neutron diffraction patterns for several
 temperatures. 
 The liquid-like peak corresponding to AF
 first neighbor correlations\cite{Canals01} starts to grow below 100 K. Below 2 K, it narrows and slightly shifts
 and an intense magnetic signal appears
 at low q values. This shows the onset of
 ferromagnetic correlations, which progressively develop as the temperature
 decreases. 
Below
 1.2 K, a magnetic contribution starts to appear on the Bragg peaks of the
 face centered cubic (fcc) lattice, which steeply increases at 0.87(2) K. This shows the onset of an
 ordered magnetic phase with a
 propagation vector {\bf k}=0.

  Rietveld refinements of the magnetic diffraction patterns (Fig. 2) were performed with
  Fullprof\cite{Carvajal93}. The magnetic structure
 was solved by a systematic search,
 using the program BasIreps\cite{BASIREPS} and symmetry-representation
 analysis\cite{Izyumov91}. 
 The basis states
 describing the Tb$^{3+}$ magnetic moments were identified and the symmetry allowed structures
 were compared to experiment. Neither a collinear ferromagnetic
 structure
 nor the {\bf k}=0 AF structures allowed by Fd$\overline{3}$m symmetry
were compatible with the data,
  yielding extinctions of several Bragg peaks. This suggests
  a magnetic component
  breaking the Fd$\overline{3}$m symmetry. Then we searched for all
  solutions in the 
   space group I4$_1$$/$amd, the highest subgroup allowing
  F and AF components simultaneously.
   The best refinement\cite{NoteIrep} (R$_{\rm B}$=2.3\%) is shown in Fig. 2 and
   Table 1.
\begin{figure} [h]
\includegraphics* [width=\columnwidth] {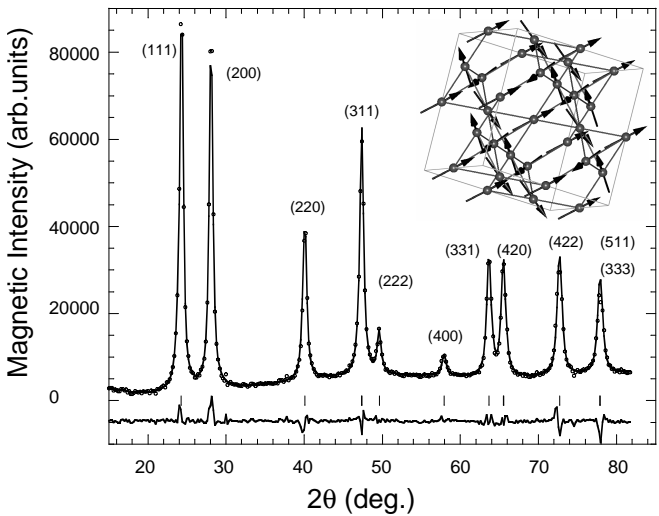}
\caption{Magnetic diffraction pattern of Tb$_2$Sn$_2$O$_7$ at 0.10
K versus the scattering angle 2$\theta$. A spectrum at 1.2 K was
subtracted. The neutron wavelength is 2.52 \AA. Solid lines show
the best refinement and the difference spectrum (bottom). In
inset, the magnetic structure.} \label{fig2.eps}
\end{figure}

     In the ordered structure with {\bf k}=0, the four tetrahedra of
    the cubic unit cell are equivalent. In a given tetrahedron, the Tb$^{3+}$ moments make an
    angle $\alpha$=$13.3^{o}$ with the local
    $<$111$>$ anisotropy axes connecting the center to the vertices. The components along these $<$111$>$ axes are oriented in the "
two in, two out" configuration
   of the local spin-ice structure \cite{Harris97}. The
    ferromagnetic component, which represents 37\%
    of the Tb$^{3+}$ ordered moment, 
     orders in magnetic
    domains oriented along $<$100$>$ axes. 
    The perpendicular components 
    make two couples of antiparallel vectors along $<$110$>$ edge axes of the tetrahedron. The
    ordered moment (M= 5.9(1) $\mu_{\rm B}$ at 0.1 K)
    agrees with high field magnetization data (5.5 $\mu_{\rm B}$ at 50
    kOe and
2 K)\cite{Matsuhira02}. It is reduced with respect to the free ion
moment of 9 $\mu_{\rm B}$, as in
Tb$_2$Ti$_2$O$_7$\cite{Gingras00}. With increasing temperature, M
remains almost constant up to 0.6 K. Then it steeply decreases,
showing an inflexion point which coincides with the T$_{\rm c}$
value of 0.87(2)K determined from the peaks in the specific heat
and susceptibility\cite{Matsuhira02}, and finally vanishes at
1.3(1)K (Fig. 3). The magnetic correlation length L$_{\rm c}$ was
deduced from the intrinsic peak linewidth\cite{Carvajal93}.
L$_{\rm c}$ remains constant and limited to about 190 $\AA$ up to
T$_{\rm c}$ then starts to decrease above T$_{\rm c}$. The angle
$\alpha$ is constant within the experimental error.

\begin{figure} [h]
\includegraphics* [width=\columnwidth] {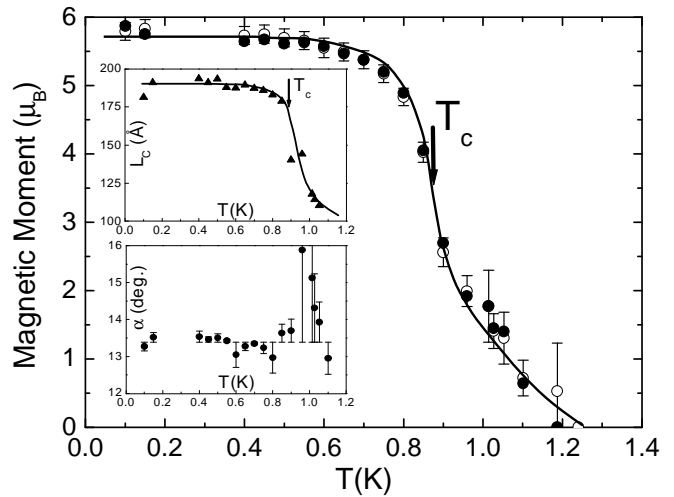}
\caption{Ordered magnetic moment M versus temperature 
(black circles) and squared intensity of the (200) magnetic peak,
scaled to the moment at 0.10 K (open circles). T$_{\rm c}$ is
determined from the peak in the specific heat. The correlation
length L$_{\rm c}$, deduced from the width of the magnetic Bragg
peaks, and the angle $\alpha$ with the local anisotropy axis are
plotted in the insets.} \label{fig3.eps}
\end{figure}

\begin{table}[!b]
\begin{center}
\begin{tabular}{ccccccc}
site & $x$ & $y$  &  $z$ &  $M_{\rm x}$ & $M_{\rm y}$ &$M_{\rm z}$ \\
\hline
1 & 0.5 & 0.5 & 0.5 & 3.85 (1) & 3.85 (1) & 2.20 (1)\\
2 & 0.25 & 0.25 & 0.5 & -3.85 (1) & -3.85 (1) & 2.20 (1)\\
3 & 0.25 & 0.5 &  0.25 & 3.85 (1) & -3.85 (1) & 2.20 (1)  \\
4 & 0.5 & 0.25 &  0.25 & -3.85 (1) & 3.85 (1) & 2.20 (1) \\
\end{tabular}
\end{center}
\caption{ \small{Magnetic components $M_{\rm x}$, $M_{\rm y}$ and
$M_{\rm z}$ of the four Tb$^{3+}$ moments (in $\mu_{B}$) in one
tetrahedron at 0.1 K. The atomic coordinates x, y, z, and the
magnetic components are expressed in the cubic unit cell.}}
\label{table1}
\end{table}

     The magnetic ground state results from a delicate balance between exchange, dipolar and anisotropy
      energies. Several theories have been developed for specific cases involving
the AF nearest neighbor exchange
     J$_{\rm nn}$, the F nearest neighbor dipolar coupling D$_{\rm nn}$ and
     the strength of the local anisotropy D$_{\rm a}$ (all taken in absolute values). A spin
liquid ground state is predicted
     for AF exchange only and Heisenberg spins\cite{Reimers91}, namely for
     J$_{\rm nn}$$>>$D$_{\rm nn}$, D$_{\rm a}$.
     The dipolar spin ice state is stabilized
       for Ising spins
      when dipolar interactions overcome the AF exchange\cite{Bramwell01,Bramwell012}, namely
     for D$_{\rm a}$$>>$D$_{\rm nn}$$>$J$_{\rm nn}$. Its local
     spin structure is similar to the observed one but spins keep the orientational disorder allowed by the "ice
     rules", 
      whereas in Tb$_2$Sn$_2$O$_7$ they order to build a {\bf k}=0 structure. When either
      a finite anisotropy\cite{Moessner98} (J$_{\rm nn}$$\geq$D$_{\rm a}$$>>$D$_{\rm nn}$)
      or a small dipolar coupling\cite{DenHertog00} (D$_{\rm a}$$>>$J$_{\rm nn}$$>$D$_{\rm nn}$) are considered, a {\bf k}=0 structure is
      predicted, but the local order differs from the present one, since in a given tetrahedron all spins point either $"$in$"$ or $"$out$"$.
      Alternatively for Heisenberg spins when dipolar
      interactions
      dominate (D$_{\rm nn}$$>$J$_{\rm nn}$$>>$D$_{\rm a}$), a {\bf k}=0 structure is predicted, but the local spin structure consists of two couples of
antiparallel moment oriented along the $<$110$>$
     edge axes of
     the tetrahedron\cite{Palmer00}. For an easy plane anisotropy (D$_{\rm a}$$<$0),
     the local order selected for the {\bf k}=0 structure, which is actually observed in Er$_2$Ti$_2$O$_7$, is also different\cite{Champion0301}. 
     The magnetic structure found in Tb$_2$Sn$_2$O$_7$ differs
     from all these ground states, and rather
      resembles the {\bf k}=0 structure of the dipolar spin ice Ho$_2$Ti$_2$O$_7$ in a low applied
      field\cite{Harris97}. It may correspond to a case not considered above, where the three parameters have
       comparable magnitudes or obey the sequence D$_{\rm nn}$$>$ D$_{\rm a}$$>$J$_{\rm nn}$.
       Its ferromagnetic character suggests that the degeneracy of the spin liquid state which results from the AF near neighbor exchange is lifted by dipolar interactions,
      the strength of the uniaxial anisotropy 
      tuning the angle $\alpha$.

The unusual change with temperature in the short range
correlations from AF to F type supports this interpretation.
Somewhat similar effects occur in the partly itinerant
LiV$_2$O$_4$, 
 or in weak ferromagnets with anisotropic interactions. 
 Here the onset of ferromagnetic
correlations just above the transition suggests that they come
from long range dipolar interactions, with
effective nearest neighbor ferromagnetic interaction\cite{Melko04}. 

 Specific heat measurements provide new information on the magnetic order in Tb$_2$Sn$_2$O$_7$.
 The temperature dependence of the specific heat C$_p$ is shown in Fig 4.
 In good agreement
 with neutron diffraction data, the specific heat C$_p$
 starts to increase below about 1.5 K
 then shows a well defined peak at 0.87
 K. 
 The final increase of C$_p$ below 0.38 K is attributed to a nuclear Schottky peak, resulting mainly from the
 splitting of the energy levels of the $^{159}$Tb nuclear spin (I=3/2) by the hyperfine field
due to the Tb$^{3+}$ electronic moment. This nuclear peak was
observed down to 0.07 K in the parent compound
Tb$_2$GaSbO$_7$\cite{Blote69}.

For T$<$0.38 K, where the nuclear contribution is dominant, the
full hyperfine hamiltonian, including a small estimated
quadrupolar term, was diagonalized  to obtain the four hyperfine
energies. Then the standard expression for a Schottky anomaly was
used to compute the nuclear specific heat $C_{nuc}$. The lines in
Fig.4 below 0.8 K represent: $C_p = C_{nuc} + C_m$, where
$C_m$=$\beta$ T$^{3}$ is an empirical electronic magnon term which
fits well the rise of $C_p$ above 0.4 K, with $\beta=12.5$ J
K$^{-4}$ mol$^{-1}$. The best fit to the data is obtained with a
hyperfine field of 135 T, which corresponds to a Tb$^{3+}$ moment
value of 3.3(3) $\mu_B$, using the hyperfine constant of 40(4)
T/$\mu_B$. The electronic entropy variation $S$ was computed by
integrating $(C_p - C_{nuc})/T$ (inset of Fig.4). In
Tb$_2$Sn$_2$O$_7$, our current measurements show that the crystal
field level scheme is only slightly modified with respect to that
in Tb$_2$Ti$_2$O$_7$, where the lowest states are two doublets
separated by 18 K \cite{Gingras00}. Therefore S should
 reach the values Rln2 and Rln4 when T increases above T$_{\rm c}$, as the first two
doublets 
 become populated. In fact, the entropy released at the
transition is only 25\% of Rln2, and it reaches 50\% of Rln2 at
1.5 K. This reflects the strong correlations of the magnetic
moments in the spin liquid phase above 1.5 K.

\begin{figure} [h]
\includegraphics* [width=\columnwidth] {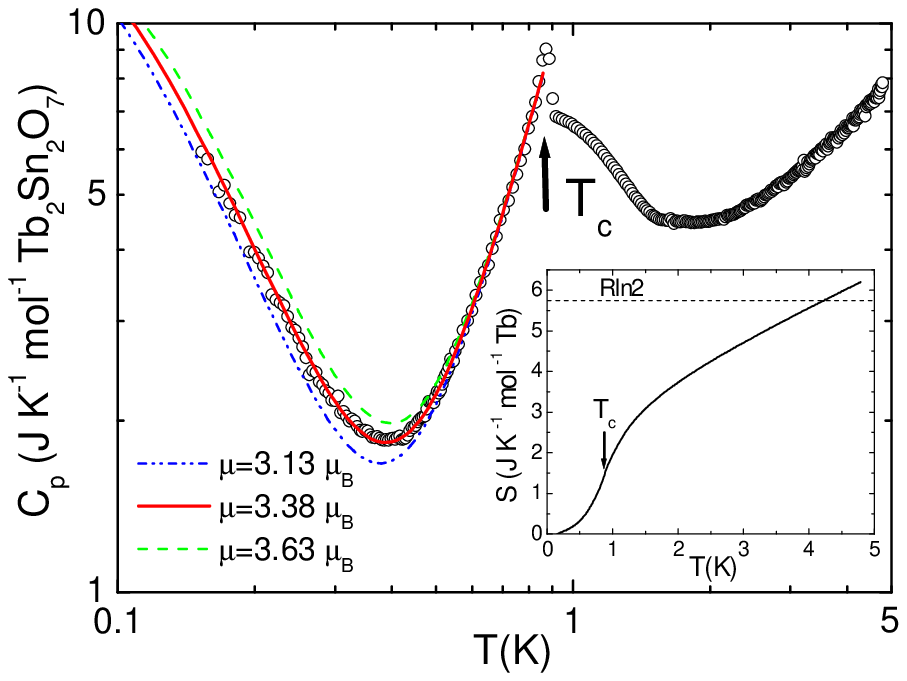}
\caption{Specific heat $C_p$ in Tb$_2$Sn$_2$O$_7$. The curves
below 0.8 K are the sum of a T$^{3}$ magnon contribution and of a
nuclear Schottky anomaly, the latter being computed for 3 moment
values (see text). The electronic entropy variation is shown in
the inset. The arrows label the transition at $T_c$=0.87 K.}
\label{fig4.eps}
\end{figure}
 The value 3.3(3) $\mu_B$ for the Tb moment deduced from the
nuclear specific heat is therefore well below the neutron value
5.9(1) $\mu_B$. Such a remarkable reduction can be explained by
the presence of electronic fluctuations. This happens if the
characteristic time $\tau$ of these
 fluctuations becomes comparable to the spin-lattice nuclear relaxation time T$_1$
 which governs 
  the thermalization of the  nuclear energy
 levels, as in Gd$_2$Sn$_2$O$_7$\cite{Bertin02}. Within the model\cite{Bertin02},
 we find that a ratio T$_1$/$\tau$= 0.4 reduces
 the nuclear specific heat by a factor 2 and accounts for
  the neutron result. This implies low temperature fluctuations of the Tb$^{3+}$ moments in the time scale $10^{-5}$ s, much slower than for paramagnetic spins ($10^{-11}s$),
which means that these fluctuations
  involve correlated spins, as previously noticed in geometrically frustrated magnets\cite{Bramwell01}.
  Here, they occur in magnetically ordered domains and may
  be connected with their finite size. They could be
  probed by $^{119}$Sn M\"ossbauer or muon spin relaxation experiments\cite{Bertin02}.

 Why does Tb$_2$Sn$_2$O$_7$ order and not
Tb$_2$Ti$_2$O$_7$? The weaker exchange energy in Tb$_2$Sn$_2$O$_7$
may not be the main reason. Our current crystal field study of the
two compounds
 by high resolution neutron scattering
 suggests another possibility.
 In Tb$_2$Sn$_2$O$_7$ only, we have observed 
 a small splitting (1.5 K) of a low energy excitation. It shows a
lifting of the degeneracy of a crystal field doublet, possibly due
to the higher value of the oxygen parameter which controls the
local distortion around the Tb$^{3+}$ ion. Assuming that the
spectral density of the spin fluctuations decreases with energy,
this lifting could weaken in Tb$_2$Sn$_2$O$_7$ the quantum
fluctuations responsible for the persistence of the spin liquid
state in Tb$_2$Ti$_2$O$_7$, and allow long range order to set in.

 In conclusion, we observed a new magnetic
structure in the spin liquid Tb$_2$Sn$_2$O$_7$. This unpredicted
structure with both ferro and antiferromagnetic character could be
called an 'ordered dipolar spin ice'. It arises below 1.3(1) K
with a low ordered moment and strong fluctuations. Then at
0.87 K a steep increase of the ordered moment coincides with a
peak in the specific heat. In the spin liquid phase, ferromagnetic
correlations replace antiferromagnetic ones
 below about 2 K.
 In the ground state,
 the lower Tb$^{3+}$ moment estimated from
 specific heat shows that the hyperfine levels are out of equilibrium and evidences the persistence
 of slow magnetic fluctuations of correlated spins.
 These unconventional fluctuations are reminiscent of the
 spin liquid in the ordered phase.

 We thank F. Bour\'ee for the neutron
 measurement on the diffractometer 3T2. We also thank F. Thomas and the cryogenic
 team of the ILL.

\end{document}